# Unveiling the mechanism of bulk spin-orbit torques within chemically disordered Fe$_x$Pt$_{1-x}$ single layers

Lijun Zhu*[1,2], Daniel C. Ralph[1,3], Robert A. Buhrman[1]

1. Cornell University, Ithaca, New York 14850, USA

2. State Key Laboratory of Superlattices and Microstructures, Institute of Semiconductors, Chinese Academy of Sciences, P.O. Box 912, Beijing 100083, China

3. Kavli Institute at Cornell, Ithaca, New York 14853, USA

*Email: lz442@cornell.edu

Recent discovery of spin-orbit torques (SOTs) within magnetic single-layers has attracted attention in the field of spintronics. However, it has remained elusive as to how to understand and how to tune the SOTs. Here, utilizing the single layers of chemically disordered Fe$_x$Pt$_{1-x}$, we unveil the mechanism of the "unexpected" bulk SOTs by studying their dependence on the introduction of a controlled vertical composition gradient and on temperature. We find that the bulk dampinglike SOT arises from an imbalanced internal spin current that is transversely polarized and independent of the magnetization orientation. The torque can be strong only in the presence of a vertical composition gradient and the SOT efficiency per electric field is insensitive to temperature but changes sign upon reversal of the orientation of the composition gradient, which are in analogue to behaviors of the strain. From these characteristics we conclude that the imbalanced internal spin current originates from a bulk spin Hall effect and that the associated inversion asymmetry that allows for a non-zero net torque is most likely a strain non-uniformity induced by the composition gradient. The fieldlike SOT is a relatively small bulk effect compared to the dampinglike SOT. This work points to the possibility of developing low-power single-layer SOT devices by strain engineering.

## 1. Introduction

Spin-orbit torques (SOTs) have attracted remarkable attention due to their potential for improving the efficiency of magnetic memory [1-5] and logic [6]. Most studies have examined cases in which the SOTs arise due to exchange interaction of a magnetic layer with an externally generated spin current, e.g. from an adjacent spin Hall metal [1-9] or topological surface states [10-12]. Recently, a novel bulk dampinglike SOT has been discovered at room temperature within magnetic single layers beyond a critical thickness (typically 5 nm [13]), e.g. in chemically disordered CoPt [13], amorphous CoTb [14,15], and epitaxial $L1_0$-ordered FePt [16,17]. This is surprising because a non-zero net SOT can occur only in samples which have both an imbalanced spin current and a broken inversion symmetry, the sources of which are not obvious in single layers of centrosymmetric magnetic materials. Utilization of the novel SOTs for magnetic memory and logic urgently requires understanding of the torque mechanism and the development of effective control strategies. So far, the mechanisms and the characteristics of these bulk SOTs remain unsettled, especially the form of the associated spin current generation, the origin of inversion asymmetry, the relative strength of dampinglike and fieldlike SOTs, and the evolution with temperature. The associated inversion asymmetry shows no apparent relevance to any long-range non-uniformity for chemically disordered CoPt and CoTb samples [13-15], while the bulk SOTs in thick films of epitaxial $L1_0$-FePt was accompanied by a vertical composition non-uniformity and a high degree of the $L1_0$ chemical ordering [16,17]. Compared to the dampinglike torque, the fieldlike torque was 6 times greater in some $L1_0$-FePt films [17], whereas it was much smaller in other magnetic single layers [13,15,16]. Investigation of the occurrence and manifestation of the bulk SOTs in different material systems and environments (e.g. temperatures) is urgently required for unveiling the mechanism and for developing more efficient materials for sing-layer SOT devices. So far, there has been no report on disordered FePt or on the temperature profile of the bulk SOTs.

Here, we report the observation and the characteristics of the bulk SOT that can occur in chemically disordered single layers of Fe$_x$Pt$_{1-x}$. By investigating how the SOT depends on the introduction of a controlled vertical composition gradient and on temperature, we gain insights about the mechanism of the bulk SOT. We find that, in the presence of an artificial composition gradient, the chemically disordered Fe$_x$Pt$_{1-x}$ possesses a dampinglike SOT efficiency per current density ($\xi_{DL}^j$) comparable in magnitude to that provided by a clean-limit Pt [18]. However, $\xi_{DL}^j$ vanishes for uniformly grown Fe$_{0.5}$Pt$_{0.5}$. Our measurements also indicate that the bulk torque arises from an internal, transversely-polarized effective spin current and is dominantly a dampinglike torque. The dampinglike SOT efficiency per electric field ($\xi_{DL}^E$) is essentially independent of temperature, leading us to suggest that the dominant effects of inversion asymmetry arise from strain gradients.

## 2. Samples and characterizations

For this study, we sputter-deposited two 16 nm-thick in-plane magnetized FePt films with strong bulk spin-orbit coupling: a uniform Fe$_{50}$Pt$_{50}$ layer and a Fe$_x$Pt$_{1-x}$ layer with $x$ linearly varying from 0.75 at the bottom to 0.25 at the top (see the schematic depicts in **Figure 1**a). The continuous tuning of $x$ for the Fe$_x$Pt$_{1-x}$ layer was achieved by varying the composition of each 0.2 nm subatomic layer during the film growth. In addition, we made control samples consisting of a 16 nm uniform $Fe_{0.25}Pt_{0.75}$ sample and a 16 nm uniform $Fe_{0.75}Pt_{0.25}$ sample. We chose the layer thickness of 16 nm because at this thickness the bulk dampinglike SOT in chemically disordered CoPt has saturated to its bulk strength per thickness [13]. Each sample was grown on oxidized Si substrates and capped by a 2 nm MgO layer and a 1.5 nm Ta layer that was fully oxidized upon exposure to atmosphere. No thermal annealing was performed on these samples to avoid undeliberate atomic diffusion and composition non-uniformity.

As indicated by the high-angle x-ray diffraction $\theta$-$2\theta$ patterns ($2\theta > 20°$ in Fig. 1b), both the uniform and composition-gradient samples have a chemically disordered structure, with a preferred face-centered cubic (fcc) (111) orientation along the film normal. The small-angle x-ray reflection curve ($2\theta < 10°$, Fig. 1b) decays faster for the $Fe_xPt_{1-x}$ sample than for the uniform $Fe_{50}Pt_{50}$ sample, which is consistent with reduced uniformity (e.g. in density, optical parameters, and perhaps interface smoothness) in the $Fe_xPt_{1-x}$ sample due to the artificial composition gradient. Figure 1c shows the in-plane magnetization hysteresis of each sample at 300 K measured by vibrating sample magnetometry upon sweeping the magnetic field ($H_{xy}$) within the film plane. The sharp in-plane field-induced switching reveals a well-defined in-plane anisotropy for both samples. The samples were subsequently patterned into 5×60 μm² Hall bars for determination of the dampinglike and the fieldlike SOT efficiencies by harmonic Hall voltage response (HHVR) measurements under a sinusoidal electric bias field of $E$ = 33.3 kV/m (more details of the technique can be found in previous papers [19,20]).

For an in-plane magnetized system, the dependence of the out-of-phase second harmonic Hall voltage ($V_{2\omega}$) on the angle ($\varphi$) of the in-plane field with respect to the current is given by [19,20]

$$V_{2\omega} = (V_{DL} + V_{ANE})\cos\varphi + V_{FL}\cos\varphi\cos2\varphi, \quad (1)$$

where $V_{DL} = -V_{AH}H_{DL}/2(H_{xy}+H_k)$, and $V_{FL} = -V_{PH}(H_{FL}+H_{Oe})/H_{xy}$, with $H_{DL(FL)}$ being dampinglike (fieldlike) SOT fields, $V_{AH}$ the anomalous Hall voltage, $V_{ANE}$ the anomalous Nernst voltage, $H_k$ is the effective magnetic anisotropy field that can be determined from the saturation field of the magnetic hard axis, $V_{PH}$ the planar Hall voltage, and $H_{Oe}$ the Oersted field. We determined the values of $V_{DL}+V_{ANE}$ and $V_{FL}$ for each magntitude of $H_{xy}$ by fitting the $V_{2\omega}$ data as a function of $\varphi$ to Eq. (1) (see Figure 2a). The linear fits of $V_{DL}$ versus $-V_{AH}/2(H_{xy}+H_k)$ and $V_{FL}$ versus $-V_{PH}/H_{xy}$ (see Fig. 2b,c) yield the values of $H_{DL}$ and $H_{FL}$. Using the values of $H_{DL(FL)}$, the dampinglike (fieldlike) SOT efficiencies per applied electric field can be determined as

$$\xi^E_{DL(FL)} = (2e/\hbar)\mu_0 M_s t H_{DL(FL)}/E, \quad (2)$$

where $e$ is the elementary charge, $\hbar$ the reduced Planck's constant, $\mu_0$ the permeability of vacuum, $M_s$ the average saturation magnetization of the spin-current detector, and $t$ the layer thickness of the spin-current detector. Correspondingly, the SOT efficiencies per unit current density are

$$\xi^j_{DL(FL)} = (2e/\hbar)\mu_0 M_s t H_{DL(FL)}/j_e, \quad (3)$$

where $j_e = E/\rho_{xx}$ is the charge current density and $\rho_{xx}$ is the average resistivity of the spin current generator. At 300 K, $\rho_{xx}$ = 90 μΩ cm and $M_s$ = 760 emu/cm³ for the $Fe_{50}Pt_{50}$ sample; $\rho_{xx}$ = 95 μΩ cm and $M_s$ = 550 emu/cm³ for the composition-gradient $Fe_xPt_{1-x}$ sample.

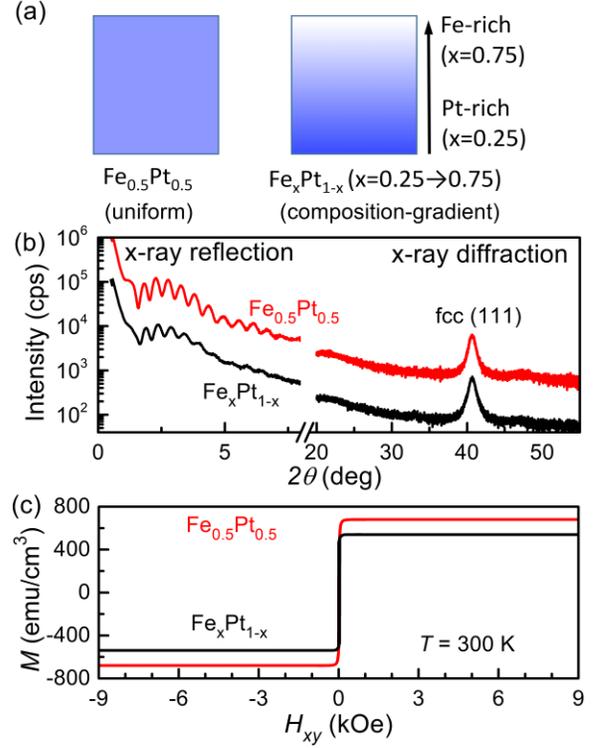

Fig. 1. Sample characterizations. (a) Schematic depictions of composition in the uniform $Fe_{50}Pt_{50}$ sample and the $Fe_xPt_{1-x}$ sample with an intentional composition gradient. (b) X-ray diffraction and reflection patterns and (c) In-plane magnetization ($M$) at 300 K versus in-plane magnetic field ($H_{xy}$) for the two types of samples.

## 3. Results and Discussion
### 3.1 Characteristics of the Spin-orbit torques

As we show in Fig. 2d, the composition-gradient $Fe_xPt_{1-x}$ sample shows a significant dampinglike torque at 300 K, with $\xi^E_{DL} \approx (-0.82 \pm 0.01) \times 10^5$ Ω⁻¹ m⁻¹ and $\xi^j_{DL} \approx -0.077 \pm 0.001$. The latter is comparable to that provided by a clean-limit Pt ($\xi^j_{DL} \approx 0.07$, $\rho_{xx}$ = 20 μΩ cm [18]) in magnitude but is of opposite sign. In contrast, $\xi^j_{DL}$ for the uniformly grown $Fe_{0.5}Pt_{0.5}$ is zero within the experimental uncertainty. The presence of a strong dampinglike torque in the $Fe_xPt_{1-x}$ sample is reaffirmed by the well-defiend spin-torque ferromagnetic resonance (ST-FMR) spectra (e.g., Fig. 2c), which contain both the symmetric and anti-symmetric components due to dampinglike and fieldlike torques [18] (we do not attempt a quantitative analysis of the ST-FMR data because this would require a detailed understanding of the distribution of non-uniform current density in the presence of the composition gradient). The negligible SOTs of the uniform $Fe_{0.5}Pt_{0.5}$ sample are also reaffirmed by the absence of any resonance signal in its ST-FMR spectrum (Fig. 2f). These observations reveal that a composition gradient is critical for the generation of a nonzero bulk dampinglike SOT in the chemically disordered $Fe_xPt_{1-x}$. This is interesting and even suprising because a composition gradient has been reported not to be necessary for the bulk SOT in chemically disordered CoPt and CoTb [13-15]. Future theoretical calculations could be informative for understanding the microscopic differences between these materials.

Figure 2d also plots the effective efficiency of the fieldlike torque, $\xi_{FL}^j$, for both samples. $\xi_{FL}^j$ for the uniform $Fe_{0.5}Pt_{0.5}$ is negligible, which is consistent with the negligible damping-like torque and net Oersted field. The positive room-temperature $\xi_{FL}^j$ value for the composition-gradient $Fe_xPt_{1-x}$ sample contains a positive contribution from the Oersted field torque due to the magnetization non-uniformity within the $Fe_xPt_{1-x}$ (the $Fe_{0.25}Pt_{0.75}$ at the bottom of the sample has the same resistivity but much lower saturation magnetization than the $Fe_{0.75}Pt_{0.25}$ at the top at 300 K, see below; the positive direction for Oersted field and fieldlike SOT effective field due to a current flow in the $+x$ direction is defined as the $-y$ direction in the measurements) and is thus higher than the true value of the fieldlike SOT. Therefore, $\xi_{FL}^j$ is, at least, more than a factor of 2 smaller than the dampinglike torque, which is reminiscent of the cases of chemically disordered CoPt [13], amorphous CoTb [15], $L1_0$-FePt [16], and the spin Hall metal/FM bilayers [7,19]. However, this is in contrast to a previous report of epitaxial $L1_0$-FePt with a composition gradient induced by annealing of up to 760 °C [17]. In that case, the fieldlike torque was reported to be five times greater than the dampinglike torque, which might be partly because the so-called "planar Hall correction" which is found to cause errors (see the Supplementary Materials of [21,22]) was applied in their "out-of-plane" HHVR analysis.

### 3.2 Source of the spin current

The occurrence of the non-zero dampinglike SOT in the composition-gradient $Fe_xPt_{1-x}$ sample indicates a non-equilibrium spin population (accumulation) in the bulk of the $Fe_xPt_{1-x}$ single layer, more precisely, the combination of an internal spin current and broken mirror and rotational symmetries. Note that in a magnetic layer that has either a mirror symmetry about its midplane or a two-fold rotational symmetry about the current axis, spin accumulation due to the spin Hall effect (SHE) and its exchange interaction with the magnetization must be equal and opposite on the two sides of the midplane [23]. Since the angle-dependent in-plane HHVR technique is sensitive only to a SOT arising from a spin current that is transversely polarized and independent of the magnetization orientation [13], the spin current associated with the detected bulk spin torque is most likely due to a bulk SHE. The anomalous Hall effect [24,25] and the planar Hall effect [26,27], which can only generate spin current collinear with the magnetization, can be excluded as possible sources.

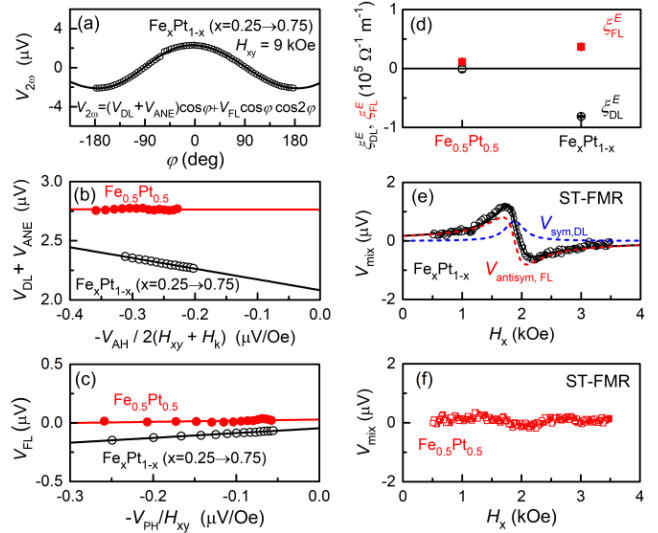

**Fig. 2**. Spin-orbit torque measurements. (a) Second harmonic Hall voltage ($V_{2\omega}$) at 300 K versus in-plane angle of the magnetization with respect to the bias current for the composition-gradient $Fe_xPt_{1-x}$ sample. (b) $V_{DL}+V_{ANE}$ versus $-V_{AH}/2(H_{xy}+H_k)$, (c) $V_{FL}$ versus $V_{PH}/H_{xy}$, (d) $\xi_{DL}^E$, $\xi_{FL}^E$ for the uniform $Fe_{0.5}Pt_{0.5}$ sample and the composition-gradient $Fe_xPt_{1-x}$ sample. (e) ST-FMR spectrum for the uniform $Fe_xPt_{1-x}$ sample showing symmetric and anti-symmetric components due to dampinglike and fieldlike torques, respectively. (f) ST-FMR spectrum for the uniform $Fe_{0.5}Pt_{0.5}$ sample, indicating the absence of any dampinglike and fieldlike torques. Both ST-FMR spectra in (e) and (f) were measured with the same rf frequency of 11 GHz and the same rf power of 5 dBm.

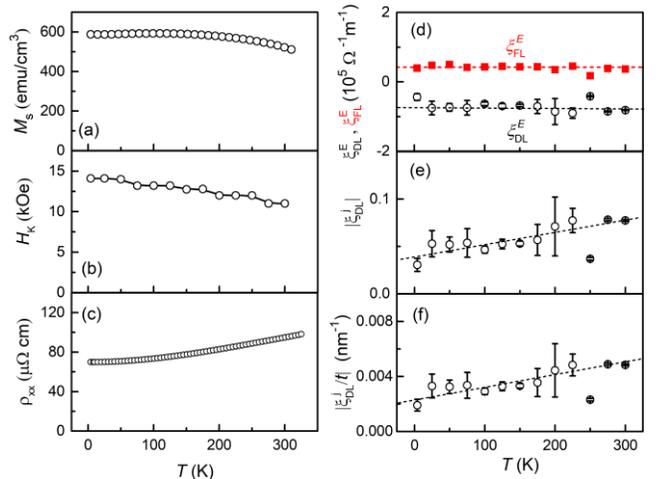

**Fig. 3**. Temperature dependence. (a) Saturation magnetization, (b) Magnetic anisotropic field, (c) Resistivity, (d) $\xi_{DL}^E$, $\xi_{FL}^E$, (e) $|\xi_{DL}^j|$, and (f) $|\xi_{DL}^j/t|$ for the 16 nm composition-gradient $Fe_xPt_{1-x}$ sample.

### 3.3 Source of the symmetry breaking

The necessity of a vertical composition gradient suggests that the long-range composition gradient is related to or enhances the asymmetry of the generation, spin relaxation, or exchange interaction of spin current in the "bulk" of the $Fe_xPt_{1-x}$. However, there still remains the question of how the composition gradient breaks the inversion symmetry. Given the previous experimental demonstration of bulk dampinglike SOT in compositionally uniform CoPt and CoTb samples [13,14], the composition gradient itself or the consequent bulk spin-orbit coupling gradient is unlikely to be the direct mechanism for the non-zero SOT in our $Fe_xPt_{1-x}$. Instead, we surmise that the more relevant non-uniformities should occur inherently during growth or be induced/enhanced by the composition gradient, e.g. non-uniformities of the magnetic moment density, the electron scattering strength, the exchange stiffness, and/or the lattice strain.

Next we show that the measurements of the SOTs as a function of temperature ($T$) can shed more insight into the underlying mechanism. We first carried out temperature-dependent HHVR measurement on the composition-gradient $Fe_xPt_{1-x}$ sample. As shown in **Figure 3**a-d, despite the significant increases upon cooling in the average values of $M_s$ and $H_k$ and the reduction in the average value of $\rho_{xx}$, there is minimal temperature dependence in both $\xi_{DL}^E$ and $\xi_{FL}^E$ (note that the moderate decrease of $\xi_{DL}^j$ upon cooling in Fig. 3e is solely due to the reduction of $\rho_{xx}$, as shown in Fig. 3c).

This first suggests that the mechanism by which the composition gradient of the $Fe_xPt_{1-x}$ breaks the inversion symmetry is not through affecting the local magnetic moment density, or local electron scattering strength, or local exchange stiffness strength. This is because these properties of the Fe-rich and Pt-rich regions of the $Fe_xPt_{1-x}$ layer should vary rather differently as a function of temperature. As indicated by the measurements on the 16 nm-thick uniformly grown $Fe_{0.75}Pt_{0.25}$ and $Fe_{0.25}Pt_{0.75}$ films (**Figure 4**a,b), upon cooing from 300 K to 4 K, the gradient of the local magnetic moment density should decrease by a factor of two (the relative difference of $M_s$ decreases from 800 emu/cm$^3$ to 400 emu/cm$^3$ because of the sensitivity of the Curie temperature of $Fe_xPt_{1-x}$ to the composition $x$ [28,29]), while the gradient of the electron scattering strength should increase by a factor of 19 (the relative difference of $\rho_{xx}$ varies from -1.0 μΩ cm to 18 μΩ cm). The exchange stiffness constant is expected to vary more rapidly than $M_s$ as a function of temperature [30] and thus is even less likely to explain the symmetry breaking.

After we have excluded non-uniformities in magnetic moment density, electron scattering strength, and exchange stiffness as the dominant mechanisms of inversion asymmetry, the lattice strain gradient is left as the most-likely symmetry-breaking mechanism for the bulk dampinglike SOT. Microscopically, a strain gradient can non-uniformly modify the strengths of the SHE [31-35], spin-orbit interaction [36], orbital polarization [37], spin states at the Fermi level [37], and strain-spin coupling [38], ultimately leading to inversion symmetry breaking in the generation and relaxation of spin current within the sample. As shown in Fig. 4c, the lattice constant of the chemically disordered $Fe_xPt_{1-x}$ indeed increases by 2.3% as $x$ from 0.25 to 0.75 [39], suggesting a very strong strain gradient in the composition-gradient samples. Meanwhile, the lattice parameter of $Fe_xPt_{1-x}$ alloys has been established to be essentially temperature-independent (changing by a factor of <5×10$^{-5}$) between 300 K and 4.2 K [40]. Therefore, we conclude that the strain is most likely the mechanism by which the composition gradient of the $Fe_xPt_{1-x}$ introduces the symmetry breaking necessary to generate a dampinglike SOT. We also find that, when the orientation of the composition gradient of the $Fe_xPt_{1-x}$ is reversed, the dampinglike and fieldlike torques reverse their signs without any significant change in the magnitudes (see Figure S1 in the Supporting Information), which strongly supports our conclusion that the dominant effects of inversion asymmetry arise from strain gradients. The mechanism could also explain the torque in epitaxial $L1_0$-FePt [17] and amorphous GdFeCo with an obvious composition gradient [41]. This type of unintentional strain gradient may also be the cause of bulk SOTs in previous experiments on CoPt and CoTb [13-15] with no intentional composition gradient. This is because such a strain gradient could arise simply from the accumulation and relaxation of strain set by the substrate in other thick films even without a deliberate composition gradient and can be challenging to detect in single layers by commonly used transmission electron microscopy or x-ray diffraction.

The robustness of $\xi_{DL}^E$ against temperature is consistent with spin-current generation by a bulk SHE, which has been found to yield $\xi_{DL}^E$ or effective spin Hall conductivity that is insensitive to temperature in HM/FM systems (including those with a dominant intrinsic SHE mechanism)[42-44]. The fieldlike torque from interfaces of HM/FM/oxide samples, however, does decrease strongly with decreasing temperature [44]. Therefore, the temperature insensitivity of the fieldlike torque in the composition-gradient $Fe_xPt_{1-x}$ sample suggests a bulk rather than interfacial origin.

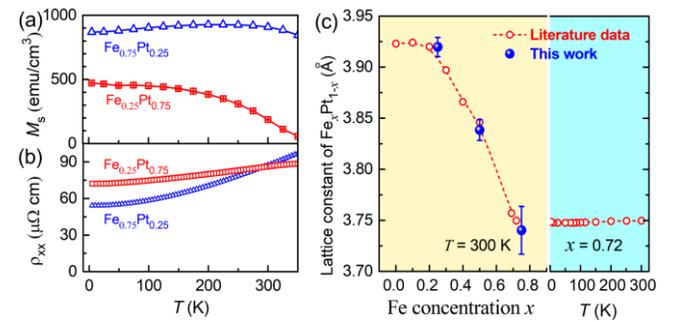

**Fig. 4**. Critical role of strain. (a) Saturation magnetization vs temperature and (b) Resistivity vs temperature for uniform-composition $Fe_{0.75}Pt_{0.25}$ and $Fe_{0.25}Pt_{0.75}$ films. (c) Lattice constant vs the Fe concentration $x$ (left) and lattice constant vs temperature (right) of uniformly grown $Fe_xPt_{1-x}$ films with a disordered fcc structure. In (c), the red circles plot the literature values of the lattice constant from Ref. [39] ($x$ < 0.5) and Ref. [40] ($x$ = 0.68 and 0.72).

### 3.4 Technological impact

Finally, we discuss the efficiency of the bulk dampinglike torque from technological point of view. To make a useful comparison of the current efficiencies, we consider the dampinglike SOT efficiency per current density per spin-current-generator thickness, $\xi_{DL}^j/t$, as an indicator for the strength of such bulk torques. Note that $\xi_{DL}^j/t$ better describes the current efficiency than $\xi_{DL}^j$ because the bulk SOT occurs only when the magnetic layer is very thick and the required current is the product of the current density and the layer thickness. This is different from the SOT in HM/FM heterostructures where the spin current is generated by the HM and the torque efficiencies are essentially independent of the layer thickness of the spin-current detector (i.e. the FM)[20,22]. Given $\xi_{DL}^E$ = -0.82×10$^5$ Ω$^{-1}$ m$^{-1}$ and $\xi_{DL}^j$ = -0.08 for the 16 nm composition-gradient Fe$_x$Pt$_{1-x}$ sample, the value of $\xi_{DL}^j/t$ is less than 0.005 nm$^{-1}$ (Fig. 3f), which is more than 10 times smaller compared to Pt 4 nm/FM bilayers (≈ 0.055 nm$^{-1}$, FM = Co, FeCoB, Ni$_{81}$Fe$_{19}$)[45]. Moreover, the bulk torque has to drive the thick magnetic layer that generates the torques, while the spin torque of the HM/FM only needs to drive a very thin magnetic layer, e.g. 1~2 nm in the three-terminal nonvolatile SOT-driven magnetic memories[3-5]. Therefore, in term of power and current requirements, the bulk dampinglike SOT is much less efficient than that in Pt-based HM/FM bilayers [20,22,45]. This fact is not specific to our Fe$_x$Pt$_{1-x}$ sample but rather universally true for the various reported magnetic single layers (e.g. 0.0075 nm$^{-1}$ for CoPt[13] and 0.016 nm$^{-1}$ for GdFeCo[41]). However, it is possible that larger gradients of strain might enable more efficient bulk SOTs in thinner single magnetic layers, which requires future efforts and is beyond of scope of this work.

### 4. Conclusion

We have measured the characteristics of bulk SOTs in chemically disordered Fe$_x$Pt$_{1-x}$ single layers with and without an intentionally-induced composition gradient. We find that, with a strong composition gradient, the Fe$_x$Pt$_{1-x}$ produces a dampinglike torque efficiency $\xi_{DL}^j$ of -0.08, close in magnitude to that provided by clean-limit Pt [18]. In contrast, $\xi_{DL}^j$ is vanishingly small for the Fe$_{50}$Pt$_{50}$ with no intentional composition gradient. We also find that the bulk torque is dominantly a dampinglike torque that arises from an internal, transversely-polarized spin current, suggesting that the associated spin current is most likely from a bulk SHE. The dampinglike SOT efficiency per electric field is insensitive to temperature, from which we argue that the inversion asymmetry necessary for the bulk SOT to exist is most likely a strain non-uniformity induced by the composition-gradient. These results provide important information for a unified understanding of the "unexpected" bulk spin-orbit torques in various magnetic single layers. Our finding also suggests that larger gradients of strain might enable more efficient bulk SOTs, and perhaps stronger bulk torques in thinner single magnetic layers, which would be necessary for this effect to be useful for applications.

### Experimental Section

*Sample preparation:* The samples were deposited at room temperature by sputtering onto oxidized Si substrates with an argon pressure of 2 mTorr and a base pressure of ~10$^{-9}$ Torr. Each sample was capped by a MgO 2 nm/Ta 1.5 nm bilayer that was fully oxidized upon exposure to the atmosphere. The samples were patterned by photolithography and ion milling into 5×60 μm$^2$ Hall bars and 10×20 μm$^2$ microstrips, followed by deposition of 5 nm Ti and 150 nm Pt as electrical contacts for harmonic Hall voltage response measurements and for ST-FMR measurements.

*Measurements*: A Rigaku Smartlab diffractometer was used for x-ray diffraction measurements. The saturation magnetization of each sample was measured at 300 K with a vibrating sample magnetometer (sensitivity ~10$^{-7}$ emu). Anomalous Hall voltage and effective anisotropic field were measured electrically using a Quantum Design physical property measurement system (PPMS). During the harmonic Hall voltage response measurements, a Signal Recovery DSP Lock-in Amplifier (Model 7625) was used to source a sinusoidal voltage onto the Hall bars and to detect the first and second harmonic Hall voltage responses. For the spin-torque ferromagnetic resonance measurements, a rf signal generator and a Signal Recovery DSP Lock-in Amplifier (Model 7625) were used and an in-plane magnetic field was swept at 45º with respect to the magnetic microstrip. An electromagnet with a maximum in-plane field of 3.5 kOe was used during the spin torque measurements.

### Acknowledgments

This work was supported in part by the Office of Naval Research (N00014-15-1-2449), in part by the NSF MRSEC program (DMR-1719875) through the Cornell Center for Materials Research, and in part by the NSF (ECCS-1542081) through use of the Cornell Nanofabrication Facility/National Nanotechnology Coordinated Infrastructure.

### Conflict of interest

The authors declare that they have no conflicts of interest.

### References

[1] L. Liu, C.-F. Pai, Y. Li, H. W. Tseng, D. C. Ralph, R. A. Buhrman, Science **2012**, 336, 555.
[2] S. Fukami, T. Anekawa, C. Zhang, H. Ohno, Nat. Nanotech. **2016**, 11, 621–625.
[3] L. Zhu, L. Zhu, S. Shi, D. C. Ralph, R. A. Buhrman, Adv. Electron. Mater. **2020**, 6, 1901131.
[4] E. Grimaldi, V. Krizakova, G. Sala, F. Yasin, S. Couet, G. S. Kar, K. Garello, P. Gambardella, Nat. Nanotech. **2020**, 15, 111-117.
[5] M. Wang, W. Cai, D. Zhu, Z. Wang, J. Kan, Z. Zhao, K. Cao, Z. Wang, Y. Zhang, T. Zhang, C. Park, J.-P. Wang, A. Fert, W. Zhao, Nat. Electron. **2018**, 1, 582.
[6] S. C. Baek, K.-W. Park, D.-S. Kil, Y. Jang, J. Park, K.-J. Lee, B.-G. Park, Nat. Electron. **2018**, 1, 398–403.

[7] L. Zhu, L. Zhu, M. Sui, D. C. Ralph, R. A. Buhrman, Sci. Adv. **2019**, 5, eaav8025.
[8] Z. Chi, Y.-C. Lau, X. Xu, T. Ohkubo, K. Hono, M. Hayashi, Sci. Adv. **2020**, 6, eaay2324.
[9] H. Xu, J. Wei, H. Zhou, J. Feng, T. Xu, H. Du, C. He, Y. Huang, J. Zhang, Y. Liu, H.-C. Wu, C. Guo, X. Wang, Y. Guang, H. Wei, Y. Peng, W. Jiang, G. Yu, X. Han, Adv. Mater. **2020**, 32, 2000513.
[10] A. R. Mellnik, J. S. Lee, A. Richardella, J. L. Grab, P. J. Mintun, M. H. Fischer, A. Vaezi, A. Manchon, E.-A. Kim, N. Samarth, D. C. Ralph, Nature **2014**, 511, 449–451.
[11] Y. Fan, P. Upadhyaya, X. Kou, M. Lang, S. Takei, Z. Wang, J. Tang, L. He, L.-T. Chang, M. Montazeri, G. Yu, W. Jiang, T. Nie, R. N. Schwartz, Y. Tserkovnyak, K. L. Wang, Nat. Mater. **2014**, 13, 699–704.
[12] P. Li, J. Kally, S. S.-L. Zhang, T. Pillsbury, J. Ding, G. Csaba, J. Ding, J. S. Jiang, Y. Liu, R. Sinclair, C. Bi, A. DeMann, G. Rimal, W. Zhang, S. B. Field, J. Tang, W. Wang, O. G. Heinonen, V. Novosad, A. Hoffmann, N. Samarth, M. Wu, Sci. Adv. **2019**, 5, eaaw3415.
[13] L. Zhu, X. S. Zhang, D. A. Muller, D. C. Ralph, R. A. Buhrman, Adv. Funct. Mater. **2020**, 30, 2005201.
[14] R. Q. Zhang, L. Y. Liao, X. Z. Chen, T. Xu, L. Cai, M. H. Guo, H. Bai, L. Sun, F. H. Xue, J. Su, X. Wang, C. H. Wan, Hua Bai, Y. X. Song, R. Y. Chen, N. Chen, W. J. Jiang, X. F. Kou, J. W. Cai, H. Q. Wu, F. Pan, C. Song, Phys. Rev. B **2020**,101, 214418.
[15] J. W. Lee, J. Y. Park, J. M. Yuk, B.-G. Park, Phys. Rev. Appl. **2020**, 13, 044030.
[16] L. Liu, J. Yu, R. González-Hernández, C. Li, J. Deng, W. Lin, C. Zhou, T. Zhou, J. Zhou, H. Wang, R. Guo, H. Y. Yoong, G. M. Chow, X. Han, B. Dupé, J. Železný, J. Sinova, J. Chen, Phys. Rev. B **2020**, 101, 220402(R).
[17] M. Tang, K. Shen, S. Xu, H. Yang, S. Hu, W. Lü, C. Li, M. Li, Z. Yuan, S. J. Pennycook, K. Xia, A. Manchon, S. Zhou, X. Qiu, Adv. Mater. **2020**, 32, 2002607.
[18] L. Liu, T. Moriyama, D. C. Ralph, R. A. Buhrman, Phys. Rev. Lett. **2011**, 106, 036601.
[19] C. O. Avci, K. Garello, M. Gabureac, A. Ghosh, A. Fuhrer, S. F. Alvarado, P. Gambardella, Phys. Rev. B **2014**, 90, 224427.
[20] L. Zhu, D. C. Ralph, R. A. Buhrman, Phys. Rev. Appl. **2018**, 10, 031001.
[21] J. Torrejon, J. Kim, J. Sinha, S. Mitani, M. Hayashi, M. Yamanouchi, H. Ohno, Nat. Commun. **2014**, 5, 4655.
[22] L. J. Zhu, K. Sobotkiewich, X. Ma, X. Li, D. C. Ralph, R. A. Buhrman, Adv. Funct. Mater. **2019**, 29, 1805822.
[23] A. Davidson, V. P. Amin, W. S. Aljuaid, P. M. Haney, X. Fan, Phys. Lett. A **2020**, 384, 126228.
[24] Tomohiro Taniguchi, J. Grollier, M. D. Stiles, Phys. Rev. Appl. **2015**, 3, 044001.
[25] J. D. Gibbons, D. MacNeill, R. A. Buhrman, D. C. Ralph, Phys. Rev. Applied **2018**, 9, 064033.
[26] C. Safranski, E. A. Montoya, I. N. Krivorotov, Nat. Nanotech. **2019**, 14, 27–30.
[27] C. Safranski, J. Z. Sun, J.-W. Xu, A. D. Kent, Phys. Rev. Lett. **2020**, 124, 197204.
[28] C.-B. Rong, Y. Li, J. P. Liu, J. Appl. Phys. **2007**, 101, 09K505.
[29] Y. Ou, D. C. Ralph, R.A. Buhrman, Phys. Rev. Lett. **2018**,120, 097203.
[30] U. Atxitia, D. Hinzke, O. Chubykalo-Fesenko, U. Nowak, H. Kachkachi, O. N. Mryasov, R. F. Evans, R. W. Chantrell, Phys. Rev. B **2010**, 82, 134440.
[31] B. A. Bernevig, S.-C. Zhang, Phys. Rev. Lett. **2006**, 96, 106802.
[32] G. Y. Guo, Y. Yao, Q. Niu, Phys. Rev. Lett. **2005**, 94, 226601.
[33] P. C. Lou, A. Katailiha, R. G. Bhardwaj, T. Bhowmick, W. P. Beyermann, R. K. Lake, S. Kumar, Phys. Rev. B **2020**, 101, 094435.
[34] J. Sławińska, F. T. Cerasoli, H. Wang, S. Postorino, A. Supka, S. Curtarolo, M. Fornari, M. B. Nardelli, 2D Mater. **2019**, 6, 025012.
[35] Y. Araki, Sci. Rep. **2018**, 8, 15236.
[36] E. Paris, Y. Tseng, E. M. Pärschke, W. Zhang, M. H. Upton, A. Efimenko, K. Rolfs, D. E. McNally, L. Maurel, M. Naamneh, M. Caputo, V. N. Strocov, Z. Wang, D. Casa, C. W. Schneider, E. Pomjakushina, K. Wohlfeld, M. Radovic, T. Schmitt, PNAS **2020**, 117 24764-24770.
[37] M. Filianina, J.-P. Hanke, K. Lee, D.-S. Han, S. Jaiswal, A. Rajan, G. Jakob, Y. Mokrousov, M Kläui, Phys. Rev. Lett. **2020**,124. 217701.
[38] D.-L. Zhang, J. Zhu, T. Qu, D. M. Lattery, R. H. Victora, X. Wang, J.-P. Wang, Sci. Adv. **2020**, 6, eabb4607.
[39] C. L. Canedy, G. Q. Gong, J. Q. Wang, G. Xiao, J. Appl. Phys. **1996**, 79, 6126-6128.
[40] F. Ono, H. Maeta, T. Kittaka, J. Magn. Magn. Mater. **1983**, 31-34, 113-114.
[41] D. Céspedes-Berrocal, H. Damas, S. Petit-Watelot, D. Maccariello, P. Tang, A. Arriola-Córdova, P. Vallobra, Y. Xu, J.-L. Bello, E. Martin, S. Migot, J. Ghanbaja, S. Zhang, M. Hehn, S. Mangin, C. Panagopoulos, V. Cros, A. Fert, J.-C. Rojas-Sánchez, Adv. Mater. **2021**, 33, 2007047.
[42] E. Sagasta, Y. Omori, M. Isasa, M. Gradhand, L. E. Hueso, Y. Niimi, Y. Otani, and F. Casanova, Phys. Rev. B **2016**, 94, 060412(R).
[43] L. Vila, T. Kimura, Y. Otani, Phys. Rev. Lett. **2017**, 99, 226604.
[44] Y. Ou, C.-F. Pai, S. Shi, D. C. Ralph, R. A. Buhrman, Phys. Rev. B **2016**, 94, 140414(R).
[45] L. Zhu, D. C. Ralph, R. A. Buhrman, Phys. Rev. Lett. **2019**, 123, 057203.

Supporting Information for

**Unveiling the mechanism of bulk spin-orbit torques within chemically disordered Fe$_x$Pt$_{1-x}$ single layers**

Lijun Zhu*[1,2], Daniel C. Ralph[1,3], Robert A. Buhrman[1]

1. Cornell University, Ithaca, New York 14850, USA
2. State Key Laboratory of Superlattices and Microstructures, Institute of Semiconductors, Chinese Academy of Sciences, P.O. Box 912, Beijing 100083, China
3. Kavli Institute at Cornell, Ithaca, New York 14853, USA

*Email: lz442@cornell.edu

**1. Harmonic Hall voltage response measurement on the Fe$_x$Pt$_{1-x}$ with reversed composition gradient**

For in-plane magnetized samples, the second harmonic Hall voltage responses to the dampinglike and fieldlike torques are given by $V_{DL} = -V_{AH}H_{DL}/2(H_{xy}+H_k)$ and $V_{FL} = - V_{PH}(H_{FL}+H_{Oe})/H_{xy}$, respectively. Here $V_{AH}$ is the anomalous Hall voltage, $H_k$ is the effective magnetic anisotropy field that can be determined from the saturation field of the magnetic hard axis, $V_{PH}$ the planar Hall voltage, and $H_{Oe}$ the Oersted field. Therefore, the values of dampinglike spin-orbit torque effective field ($H_{DL}$) and fieldlike torque effective field ($H_{FL}$) are the slopes of the linear fits of $V_{DL}$ versus $-V_{AH}/2(H_{xy}+H_k)$ and $V_{FL}$ versus $-V_{PH}/H_{xy}$. As can be clearly seen from **Figure S1**a,b, both $H_{FL}$ and $H_{FL}$ for the control sample Fe$_x$Pt$_{1-x}$ ($x = 0.75\rightarrow0.25$) shows comparable magnitude but reversed sign compared to the Fe$_x$Pt$_{1-x}$ ($x = 0.25\rightarrow0.75$) discussed in detail in the main text, while $H_{FL}$ and $H_{FL}$ for the uniform Fe$_{0.5}$Pt$_{0.5}$ sample are negligibly small compared to the composition-gradient samples. The sign change of both dampinglike and fieldlike torques of the Fe$_x$Pt$_{1-x}$ due to reversal of the direction of the composition gradient strongly supports our conclusion that the dominant effects of inversion asymmetry arise from strain gradients.

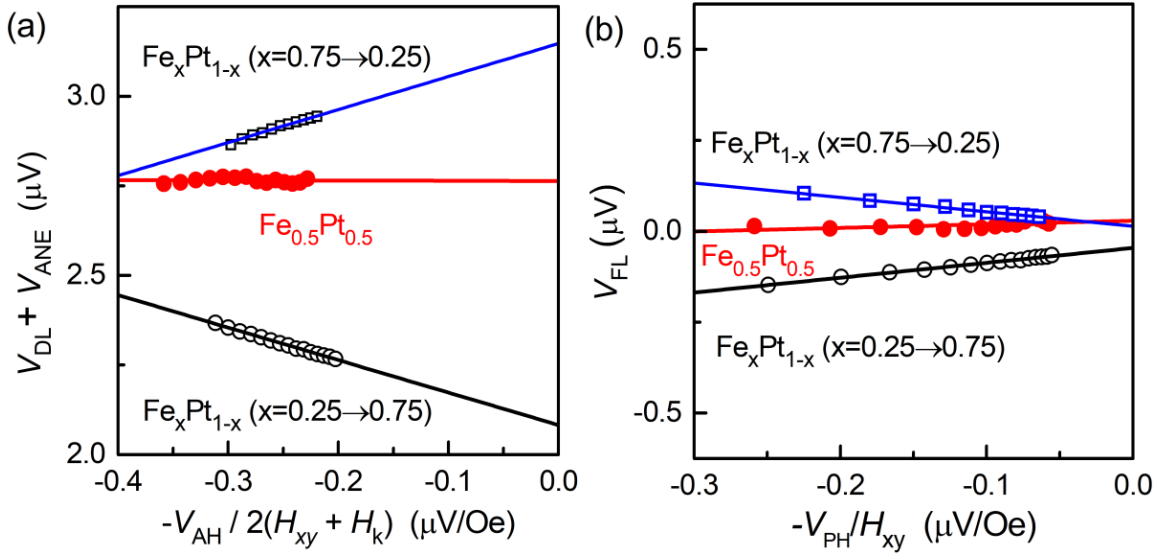

Figure S1. Sign change of the spin-orbit torque efficiencies upon reversal of the composition gradient. (a) $V_{DL}+V_{ANE}$ versus $-V_{AH}/2(H_{xy}+H_k)$ and (b) $V_{FL}$ versus $V_{PH}/H_{xy}$ for the uniform Fe$_{0.5}$Pt$_{0.5}$ sample and the composition-gradient Fe$_x$Pt$_{1-x}$ ($x = 0.75\rightarrow0.25$) and Fe$_x$Pt$_{1-x}$ ($x = 0.25\rightarrow0.75$) samples.

**2. Ferromagnetic resonance spectra of the control samples**

One might consider that the spin-torque ferromagnetic resonance (ST-FMR) spectra of the control samples may provide additional information on the role of the strain and composition gradient. To properly understand the ST-FMR results, we first measured the magnetic damping ($\alpha$) of these samples because $\alpha$ and the associated FMR linewidth ($\Delta H$) play a key role in the determination of the amplitude of FMR spectra of the samples. Using ST-FMR in the rf frequency ($f$) regime of 6-14 GHz, we measured $\alpha$ for each sample from the best fits of $\Delta H$ (half width at half maximum) to the relations $\Delta H = \Delta H_0 + 2\pi\alpha f / \gamma$, where $\Delta H_0$ is the inhomogeneous broadening of the FMR linewidth and $\gamma$ is the gyromagnetic ratio. The obtained $\alpha$ value is $0.011 \pm 0.002$ for the uniform $Fe_{0.75}Pt_{0.25}$, $0.041 \pm 0.008$ for the composition-gradient $Fe_xPt_{1-x}$ ($x = 0.25 \rightarrow 0.75$), and $0.337 \pm 0.043$ for the composition-gradient $Fe_xPt_{1-x}$ ($x = 0.75 \rightarrow 0.25$). Since the uniform $Fe_{0.5}Pt_{0.5}$ shows no detectable ST-FMR signal (see Figure 2(f) in the main text), we determine its $\alpha$ value as only $0.023 \pm 0.001$ using flip-chip FMR excited by rf current within a coplanar waveguide below the sample.

In Figure S2a-e we show the ST-FMR spectra at 11 GHz for the samples studied in this work: the uniform $Fe_{0.75}Pt_{0.25}$, the uniform $Fe_{0.25}Pt_{0.75}$, the composition-gradient $Fe_xPt_{1-x}$ ($x = 0.25 \rightarrow 0.75$), and the composition-gradient $Fe_xPt_{1-x}$ ($x = 0.75 \rightarrow 0.25$). After correction for the approximately inverse parabolic scaling of the amplitude of the FMR signal with the FMR linewidth or damping (i.e. $V_{\text{mix}} \propto 1/\Delta H^2$ or $1/\alpha^2$), we can estimate that the FMR amplitude of the uniform $Fe_{0.75}Pt_{0.25}$ is not only much weaker than composition-gradient samples $Fe_xPt_{1-x}$ ($x = 0.25 \rightarrow 0.75$) but also weaker than the $Fe_xPt_{1-x}$ ($x = 0.75 \rightarrow 0.25$). Since the harmonic Hall voltage response measurements have determined that $Fe_xPt_{1-x}$ ($x = 0.75 \rightarrow 0.25$) has a large dampinglike and fieldlike torque comparable to the $Fe_xPt_{1-x}$ ($x = 0.25 \rightarrow 0.75$), we attribute the relatively small amplitude of the ST-FMR signal of $Fe_xPt_{1-x}$ ($x = 0.75 \rightarrow 0.25$) solely to the very high damping. The latter is confirmed by the very weak flip-chip FMR spectrum shown in Figure S2e. Note that the flip-chip FMR is excited by rf current within the waveguide below the sample and is thus irrelevant to the strength of spin torque.

It is an interesting observation that the orientation reversal of the composition gradient makes the damping so different for the $Fe_xPt_{1-x}$ ($x = 0.25 \rightarrow 0.75$) and $Fe_xPt_{1-x}$ ($x = 0.75 \rightarrow 0.25$). The detailed mechanism is worth future study but is beyond the scope of this work. We also find that the uniform $Fe_{0.75}Pt_{0.25}$ with very small damping shows a small but detectable FMR signal, which suggests the occurrence of an unintentional strain gradient, e.g. from the substrate or the strain relaxation. The uniform $Fe_{0.25}Pt_{0.75}$ (Figure S2b) shows no distinguishable signal because the torque should be small and because the damping is remarkable as the samples approaches its Curie temperature of slightly higher than 300 K. The latter is confirmed by negligibly weak flip-chip FMR (Figure S2e).

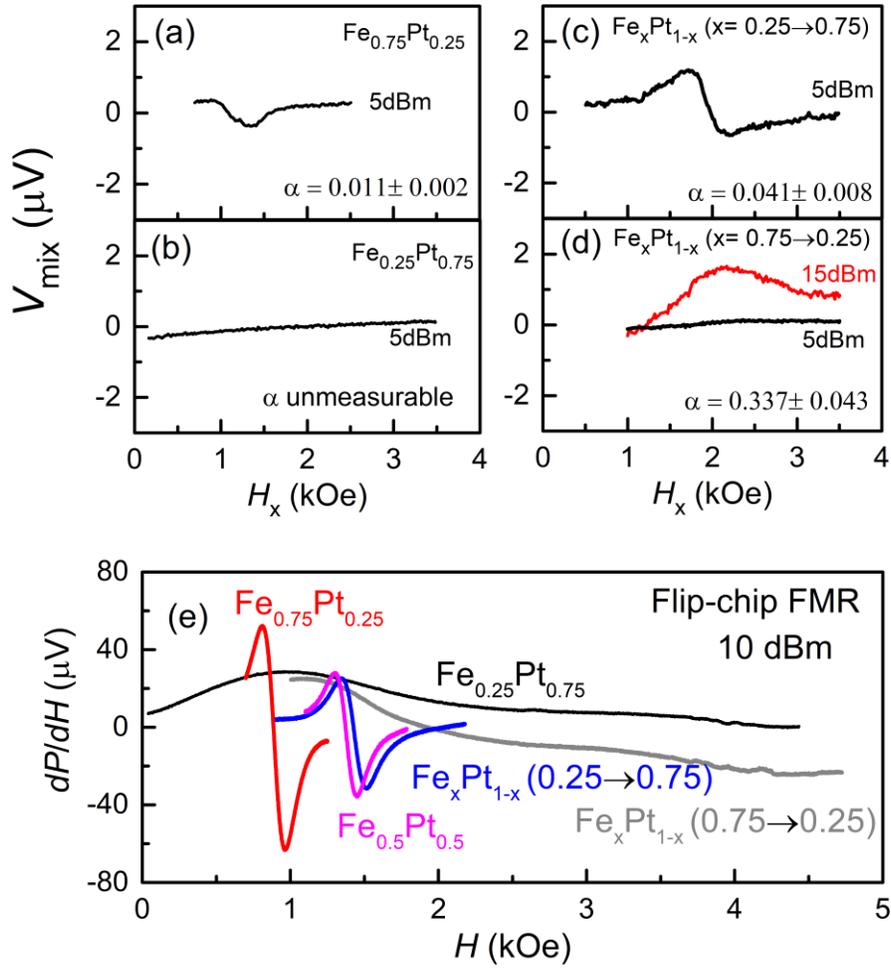

Figure S2. Room-temperature ST-FMR spectra for (a) the uniform $Fe_{0.75}Pt_{0.25}$ sample (damping $\alpha \approx 0.011$, 5dBm, 11 GHz), (b) the uniform $Fe_{0.25}Pt_{0.75}$ sample ($\alpha$ is too large to measure, 5dBm, 11 GHz), (c) $Fe_xPt_{1-x}$ ($x = 0.25 \rightarrow 0.75$) ($\alpha \approx 0.041$, 5dBm, 11 GHz), and (d) $Fe_xPt_{1-x}$ ($x = 0.75 \rightarrow 0.25$) ($\alpha \approx 0.337$, 5 or 15 dBm, 11 GHz). (e) Room-temperature flip-chip FMR spectra of the same samples. The very weak FMR signals of $Fe_xPt_{1-x}$ ($x = 0.75 \rightarrow 0.25$) and $Fe_{0.25}Pt_{0.75}$ in (b), (d), and (e) reveal remarkable damping in this two samples.

9